\documentclass[showpacs,twocolumn]{revtex4}
\usepackage[dvips]{epsfig}
\newcommand{\be}{\begin{equation}}
\newcommand{\ee}{\end{equation}}
\newcommand{\clF}{{\cal F}}
\newcommand{\hV}{\hat{ L}_{FP}}
\newcommand{\clL}{{\cal L}}
\newcommand{\dvr}{\mbox{div}\,}
\newcommand{\sgn}{\mbox{sgn}\,}
\newcommand{\tg}{\tilde{g}}
\newcommand{\tD}{\tilde{D}}
\newcommand{\tG}{\tilde{G}}

\newcommand{\bea}{\begin{eqnarray}}
\newcommand{\eea}{\end{eqnarray}}

\newcommand{\prt}{\partial}

\newcommand{\rgl}{\rangle}
\newcommand{\lgl}{\langle}

\begin{document}

\title{Negative superdiffusion due to the inhomogeneous convection }
\author{A. Iomin and E. Baskin  }

\affiliation{Department of Physics and Solid State Institute, \\
Technion, Haifa, 32000, Israel  }
\begin{abstract}
Fractional transport of particles on a comb structure in the presence of 
an inhomogeneous convection flow is studied.
The large scale asymptotics is considered. It is shown that a contaminant
spreads superdiffusively in the 
direction opposite to the convection flow. Conditions for the realization 
of this new effect is discussed in detail.
\end{abstract}

\maketitle

A realization of superdiffusion on a subdiffusive media, {\em e.g.} on a 
comb structure \cite{bas_iom}, is an example of the fractional transport
due to the inhomogeneous convection. These studies of the 
space--time evolution of an initial profile of particles in a specific 
media due to the inhomogeneous convection flow are arisen in a variety
of applications such as transport of external species (pollution) in water 
flows through porous geological formations \cite{berkowitz,hilfer}, 
problems of diffusion and reactions in porous catalysts \cite{andrade} 
and fractal physiology \cite{west}. The conditions for the
inhomogeneous convection which is responsible for the superdiffusive 
spreading  of an initial packet of particles have been found in 
\cite{bas_iom}. 
A classification of possible scenarios of the space--time evolution of a
contaminant in the presence of the inhomogeneous convection on the comb 
structure is presented there. The external forcing 
has been taken in the power--law form for the convection current 
$j_x(t,x,y)=vx^s\delta(y)G(t,x,y)$, where a distribution $G(t,x,y)$ 
describes 
the evolution of the initial profile, while $vx^s\delta(y)$ is the 
inhomogeneous convection velocity.
When $s<0$ subdiffusion is observed \cite{procaccia,klafter}. When $s>0$ 
it is superdiffusion.
The homogeneous convection with $s=0$ corresponds formally to the normal
diffusion, but the effective diffusion coefficient is determined by the 
external forcing $v$ \cite{bas_iom}. The frontier case with $s=1$ 
corresponds to the 
log--normal distribution of transport particles, where one deals with not 
a sum of independent random variables but with their multiplication
\cite{shlezinger}. 
When  $s>1$ the fractional transport corresponding to superdiffusion 
possesses specific features.
This case is the main subjective for the investigation in the present 
Letter.
We consider both the fractional transport on the comb 
structure, where the number of transporting particles is not conserved,
and the continuous time random walk (CTRW) with conservation of the total
number of particles (transporting or not).
This case of $s>1$ differs essentially from those with $s\le 1$ where 
the transporting particle move in the direction of the convection current
$j_x$ \cite{bas_iom}. Unlike our previous consideration \cite{bas_iom} in 
the present study we observed analytically that an asymptotic solution 
for the transporting particles corresponds to superdiffusion in the 
direction  opposite to the convection flow current $-j_x$. 
This is a result of an irreversible relaxation  process \cite{zaslavsky} 
which is ``inevitable'' process in the diffusion transport phenomena.

First, we will describe the superdiffusion on the comb structure due to 
the inhomogeneous convection described by the $2D$ distribution function
$G=G(t,x,y)$ and the current 
\be\label{dr1}
{\bf j}=\left(v(x,y)G-\tD\delta(y)\frac{\prt G}{\prt x}, 
-D\frac{\prt G}{\prt y}\right) \, ,
\ee
where $\tD\delta(y)$ and $D$ are the diffusion 
coefficients for the $x$ and $y$ directions correspondingly,
while the inhomogeneous convection velocity is $v(x,y)=v|x|^s\delta(y)$.

The comb model is known as a toy model for porous media
used for the exploration of low dimensional percolation clusters 
\cite{em1} and an electrophoresis process \cite{baskin2}. For $v=0$ 
subdiffusion have been observed \cite{baskin1}.
A special transport behavior on the comb structure is that the
displacement in the $x$--direction is possible only along
the structure $x$-axis, say  at $y=0$, according to Eq. (\ref{dr1}). 
Both the diffusion coefficient and the convection flow 
are highly inhomogeneous in the $y$-direction. There is also 
diffusion in the $y$--direction with a constant diffusion coefficient $D$.
Therefore the Liouville equation 
\[\frac{\prt G}{\prt t}+\dvr {\bf j}=0 \]
corresponds to the following Fokker--Planck equation
\be\label{dr2}
\frac{\prt G}{\prt t}+\hV(x) G\delta(y)-D\frac{\prt^2G}{\prt y^2}=0 \, 
\ee
with the Fokker--Planck operator of the form 
\[\hV (x)G =-\tD\frac{\prt^2 G}{\prt x^2}+
v|x|^s \frac{\prt G}{\prt x}+sv|x|^{s-1}\sgn(x)G \, . \]
The initial condition is $ G(0,x,y)=\delta(x)\delta(y)$, and the
boundary conditions on the infinities have the form
$G(t,\pm\infty,\pm\infty)=0$
and the same for the first derivatives with respect to $x$ and $y$
$G_x^{\prime}(t,\pm\infty,\pm\infty)=G_y^{\prime}(t,\pm\infty,\pm\infty)
=0$.
The function $\sgn(x)$ equals to $1$ for $x>0$ and $-1$ for the 
opposite case.
The transport of particles along the
structure $x$-axis is described by the function $G(t,x,y=0)=g(x,t)$.
It should be underlined that the tails of the distributions are 
the most interesting for applications. Therefore, we are studying here
the large scale asymptotics when $|x|\gg 1$.
In the Laplace--Fourier ($\clL\clF$) space $(p,k)$ Eq. (\ref{dr2})
is transformed to the following fractional equation for the 
function $\tg=\tg(p,k)=\clL\clF[g(t,x)]$
\be\label{dr3}
\tD k^2\tg(p,k)+ik
\frac{\prt^s\tg(p,-k)}{\prt |k|^s}+2\sqrt{Dp}\tg(p,k)=1 \, .
\ee
Here the fractional Reisz derivative is the result of the Fourier
transform \cite{zaslavsky, saichev}
\[-\frac{\prt^s\tg(\dots,-k)}{\prt |k|^s}=\clF[|x|^sg(\dots,x)]\, . \]
The large scale asymptotics $|x|\gg 1 $ corresponds to $|k|\ll 1$ in
the Fourier space. Therefore, the first term in (\ref{dr3}) can be omitted
at the condition 
\be\label{dr4}
\lim_{k\to 0 \atop p\to 0}\frac{k^2}{p^{1/2}}=0 \, . 
\ee 
This approximation depends on the form of singularity of the the 
convection velocity in the limit $x\rightarrow\infty$. 
It means that the asymptotic solutions of the 
homogeneous part of Eq (\ref{dr3}) for $|x|\gg 1$ 
depends on the exponent $s$ in the power law $|x|^s$ \cite{erdelyi,olver}. 
After performing the inverse Fourier transform, one obtains the asymptotic of 
$x\gg 1$ solution  that corresponds to the homogeneous part of Eq. 
(\ref{dr3}). It reads 
\be\label{dr5}
\tg(p,x)=\frac{1}{|x|^s}
\exp\left[\frac{2\sqrt{Dp}|x|^{1-s}\sgn(x)}{(s-1)v}\right] \, . 
\ee
This solution describes  asymptotic transport of any initial profile. To 
obtain the 
time--dependent solution one carries out the inverse Laplace transform
$g(t,x)=\clL^{-1}[\tg(p,x)]$. The necessary condition theorem needs
the negative sign of the function in the exponential in Eq. (\ref{dr5}).
It depends only on $s$ and the sign of $v$. When $s<1$,
the initial profile of particles moves in the directions of the convection 
flow, namely, in the direction of $v=|v|$ for $x>0$, and $v=-|v|$ for $x<0$. 
It is usual superdiffusive acceleration of diffusion 
due to the inhomogeneous convection. This case together with $s=1$ was 
considered in detail in \cite{bas_iom}. 

When $s>1$ the situation is much interesting and leads 
to the absolutely new effect. Indeed, for $s>1$, the necessary condition 
to perform the inverse Laplace transform is  $v=-|v|$, when
$x>0$ and $v=|v|$, when $x<0$. Hence, the inverse Laplace transform gives
\be\label{dr6}
g(t,x)=\frac{-\sgn(x)D^{1/2}|x|^{1-2s}}{v(s-1)\sqrt{\pi t^3}}
\exp\left[-\frac{Dx^{2-2s}}{v^2(s-1)^2t}\right] \, .
\ee
When $|x|\gg 1$ and $t$ is large enough to put the exponential to the 
unite, one obtains the distribution for superdiffusion of particles
\be\label{dr7}
g(t,x)\propto \frac{1}{|x|^{2s-1}\sqrt{\pi t^3}} \, .
\ee
All moments of $x$ higher than $2s-2$ are equal to the infinity.
It also should be underlined that the flux on the infinities is vanishing.
The important feature of this superdiffusion is that it occurs 
in the direction opposite to the inhomogeneous convection current.
This new phenomenon is related to the relaxation process or it is due to
diffusion, where the Kolmogorov conditions (see \cite{zaslavsky})
are necessary for the inferring of the Fokker--Plank equation (FPE). In 
the absence 
of the convection the solution of the FPE gives that at any moment
$t>0$ the particles are spread over the all $x$--axis from the minus 
infinity
to the plus infinity with the exponentially small tails. It is correct 
not only for the normal diffusion but for the subdiffusive relaxation on 
the comb structure, as well \cite{baskin1}
\[g(t,x)=\frac{\tD}{2\pi\sqrt{Dt^3}}\int_0^{\infty}
\exp\left[-\frac{x^2}{4\tD u}-\frac{Du^2}{t}\right]u^{1/2}du \, . \] 
This behavior is dominate for small $x$ even in the presence of the 
inhomogeneous convection. But for the asymptotically large $x$
the inhomogeneous convection in the direction opposite to the 
spreading  of particles changes the shape of the tail of the packet 
from the  exponential to the power law in according with Eq. (\ref{dr7}). 
We call this solution the 
negative superdiffusion  solution or the negative superdiffusion 
approximation (NSA).

The total number of transporting particles on the structure axis decreases 
with time 
\be\label{dr8}
 \lgl G\rgl=\int_{-\infty}^{\infty}g(t,x)dx=
(4s-3)/\sqrt{\pi Dt} \, .
\ee
Therefore, the distribution function (\ref{dr6}) 
describes the NSA when the number of particles  
$\lgl G\rgl$ is not conserved. The formulation of the NSA problem with 
conservation of the total number of particles is equivalent to the case 
with a continuous distribution of the delay times \cite{shlezinger}, where 
the total number of particles is described by the function
$G_1(t,x)=\int_{-\infty}^{\infty}G(t,x,y)dy$. It is simply to show from 
Eq. (\ref{dr2}) that 
\be\label{dr9}
G(t,x,y)=\clL^{-1}\left[\tg(p,x)e^{-\sqrt{p/D}|y|}\right] \, .
\ee
Taking this into account, one obtains the equation for $G_1$ by integrating
Eq. (\ref{dr2}) with respect to the $y$ variable. It reads in the Laplace 
space for $\tG_1(p,x)=\clL[G_1(t,x)]$:
\be\label{dr10}
p\tG_1(p,x)+\hV(x)\tg(p,x)=\delta(x)
\ee
It is simply to see from Eq. (\ref{dr2}) or Eq. (\ref{dr3}) that
\[\hV\tg(p,x)=\delta(x)-2\sqrt{pD}\tg(p,x) \, .\]
 Substituting this in Eq.
(\ref{dr10}), one obtains that 
\[\tg(p,x)=\frac{1}{2}\sqrt{p}{D}\tG_1(p,x)\, . \] 
 After substitution of this relation in Eq. (\ref{dr10}), the CTRW 
equation in the Laplace space is
\be\label{dr11}
p^{1/2}\tG_1+\frac{1}{2D}\hV(x)\tG_1=p^{-1/2}\delta(x) \, .
\ee
We introduce the Riemann--Liouville fractional derivatives 
(see, for example, \cite{klafter,mainardi})
$\frac{\prt^{\alpha}}{\prt t^{\alpha}}f(t)$ by means of 
the Laplace transform ($0<\alpha<1$):
\be\label{dr12}
 \clL[\frac{\prt^{\alpha}}{\prt t^{\alpha}}f(t)]=
p^{\alpha}\tilde{f}(p)-p^{1-\alpha}f(O^+) \, 
\ee
that also implies $\prt^{\alpha}[1]/\prt t^{\alpha}=0$ \cite{mainardi}.
Using this definition, we write down the CTRW equation which corresponds 
to the comb model desribed by Eq. (\ref{dr2})
\be\label{dr13}
\frac{\prt^{1/2}G_1}{\prt t^{1/2}}+\frac{1}{2D}\hV(x) G_1=0 \, .
\ee
Here the initial condition is $G_1(0,x)=\delta(x)$.
For the asymptotically large scale $x\gg 1$ ( or $x\ll -1$), we neglect the 
inhomogeneous term together with the second derivatives with respect to 
$x$ in Eq. (\ref{dr11}) to obtain the following equation 
\[v(d |x|^s\tG_1/dx)+2Dp^{1/2}\tG_1=0 \]
 with the NSA related to the CTRW by
\be\label{dr14}
\frac{1}{x^s}\exp\left[\frac{2Dp^{1/2}x^{1-s}\sgn(x)}{v(s-1)}\right] \, .
\ee

In the rest of the Letter we infer the NSA in the framework of the
Liouville--Green asymptotic solution for linear differential equations
\cite{olver}.
We show that the performed approximation due to the condition (\ref{dr4}) 
is satisfactory good and 
corresponds to the Liouville--Green (LG) approximation also called the WKB
approximation \cite{wkb}.
The CTRW equation (\ref{dr13}) in the generalized form reads
\be\label{dr15}
\frac{\prt^{\alpha}G_1}{\prt t^{\alpha}}+\frac{1}{2D}\hV(x) G_1=0 \, ,
\ee
where $0<\alpha<1$. Hence, for $x\gg 1$, we obtain
the homogeneous part (lhs) of Eq. (\ref{dr11}), where the
item $p^{1/2}$ is substituted by $p^{\alpha}$. It reads
\be\label{dr16}
-\tG_1^{\prime\prime}+\frac{v}{\tD}x^s\tG_1^{\prime}+
\frac{vs}{\tD}x^{s-1}\tG_1+p^{\alpha}\tG_1=0 \, .
\ee
The term in the first derivative is removed from the equation by the
substitution
\be\label{dr17} 
\tG_1=\exp[vx^{s+1}/2\tD(s+1)]w \, .
\ee
Thus $w^{\prime\prime}=R(x)w$, where 
\[ R(x)=\frac{v^2x^{2s}}{4\tD^2}[1+\frac{2s\tD}{v} x^{-s-1}
+\frac{8D\tD}{v^2}p^{1/2}x^{-2s}] \, .\]
The LG approximation for $w$, that satisfies to the accepted boundary 
conditions (see Eq. (\ref{dr2})), is 
\bea\label{dr18}
&w=BR^{-1/4}\exp\left[-\int R^{1/2}dx\right] \nonumber \\
&=B\sqrt{\frac{|v|}{2\tD}}\frac{1}{x^s}
\exp\left[-\frac{vx^{s+1}}{2\tD(s+1)}-
\frac{2Dp^{\alpha}x^{1-s}}{v(s-1)}\right] \, ,
\eea
where $B$ is a  constant. Analogously, we obtain the LG
solution for the negative $x\ll -1$. Therefore, taking 
$B=\sqrt{\frac{2\tD}{|v|}}$ and $\alpha=1/2$,
we obtain that Eq. (\ref{dr18}) coincides exactly with the solution 
(\ref{dr14}). It means that removing the second derivatives from
$\hV$ or, the same, the term $k^2$ in the Fourier space in the limit
$k\rightarrow 0$ corresponds to the Liouville--Green approximation
for the Fokker--Planck equation with inhomogeneous (superdiffusive)
convection. This asymptotic solution is superdiffusive transport of 
particles in  the direction opposite to the convection current, namely 
it is the NSA.


\begin{thebibliography}{99}
\bibitem{wkb} This approximation was used independently by Liouville and 
Green. In quantum mechanics this approximation is known as the
Wentzel--Kramers--Brillouin (WKB) approximation. However, we citing 
\cite{olver}, ``the contribution of these authors was not the construction
of the approximation (which was already known), but the determination of 
connection formulas for linking exponential and oscillatory LG 
approximations across a turning point on the real axis''.
\end{thebibliography}
\end{document}